\documentstyle[preprint,aps,version2,epsf]{revtex} 
\renewcommand{\baselinestretch}{1.1}
  \def\thebibliography#1{{\bf{References}}\list
 {[\arabic{enumi}]}{\settowidth\labelwidth{[#1]}\leftmargin\labelwidth
   \advance\leftmargin\labelsep
   \usecounter{enumi}}
   \def\newblock{\hskip .11em plus .33em minus -.07em}
   \sloppy
   \sfcode`\.=1000\relax}
  
\begin{document}
\input{epsf.tex}
\newcommand{\docuname}{\mbox{\bf  DISJET}}
\newcommand{\oas}{\mbox{$\mbox{O}(\alpha_{s})$}}
\newcommand{\oasz}{\mbox{$\mbox{O}(\alpha_{s}^{2})$}}
\newcommand{\as}{\mbox{$\alpha_{s}$}}
\newcommand{\asz}{\mbox{$\alpha_{s}^{2}$}}
\newcommand{\lms}{\mbox{$\Lambda_{\overline{\mbox{\tiny MS}}}$}}
\def\Lms#1{\Lambda_{\overline{MS}}^{(#1)}}
\def\asp{{\alpha_s\over\pi}}
\def\ams{\alpha_{\overline {MS}} }
\hbox to \hsize{
\hskip.5in \raise.1in\hbox{\bf University of Wisconsin - Madison}
\hfill$\vcenter{\hbox{\bf MAD/PH/820}
            }$
               }
\mbox{}
\hfill  May  1994   \\   
\vspace{.5in}
\vspace{0.5cm}
\begin{center}
  \begin{Large}
  \begin{bf}
Complete \oasz \hspace{2mm} Corrections to (2+1) Jet\\ Cross Sections
in Deep Inelastic Scattering\\[1cm]
  \end{bf}
  \end{Large}
  \vspace{0.8cm}
  \begin{large}
{T. Brodkorb$^a$ and E. Mirkes$^{b}$}\\[1cm]
  \end{large}
{\it
$^a$
Laboratoire Physique Theorique et Hautes
Energies, Universit{\'e} Paris-Sud, 91405 Orsay, France.
Address after June 1993, SAP Walldorf, Germany\\[2mm]
$^b$Physics Department, University of Wisconsin, Madison WI 53706, USA\\[1cm]
}
  {\bf Abstract}
\end{center}
\begin{quotation}
\noindent
Complete next-to leading order QCD predictions for
 (2+1) jet cross sections  and jet rates
in deep inelastic scattering (DIS) based on a new parton level
Monte Carlo program
 are presented.
All relevant helicity contributions to the total cross section are included.
Results on total  jet cross sections
as well as differential distributions in the basic kinematical variables
$x,W^2$ and $Q^2$
are shown for HERA energies and for the fixed target experiment
E665 at FERMILAB.
We study the dependence on the choices of the
renormalization scale $\mu_R$ and the factorization scale $\mu_F$
and show that the NLO results
are much less sensitive to
the variation of $\mu=\mu_F=\mu_R$
than the LO results.
The effect of an additional $p_T$ cut to our jet definition scheme
is investigated.
\end{quotation}
\thispagestyle{empty}
\newpage
\section{Introduction}
The start-up of the HERA electron-proton collider in
1992 marked
 the beginning of a new aera of experiments exploring
 Deep Inelastic Scattering of electrons and protons \cite{h1pub,zeuspub}.
One of the topics to be studied at HERA will be the deep
 inelastic ($\equiv$ high $Q^2$) production of multi jet
events\footnote{Much higher effective
 luminosity will be achieved for low $Q^2$
quasi-real photoproduction of jets. The theoretical interpretation of these
events, however, is more difficult, since there is added
 complication of
 the direct and the resolved photon contribution to the cross section,
which are hard to separate}, where good event statistics are expected
allowing for precision tests of QCD.
Multijet events in DIS are first  observed at the FERMILAB E665 experiment
\cite{e665pub}.

In this paper, we present complete results for (2+1) jet cross sections
(''+'' denotes the remnant jet)
in DIS based on a new parton level
Monte Carlo program \docuname\ \cite{disjet}.
All helicity contributions to the total cross section
(i.e. $\sigma_{U+L}[(2+1)-\mbox{jet}]$ and
$\sigma_L[(2+1)-\mbox{jet}]$; see below) are included.
Note, that in contrast to jet production in $e^+e^-$ experiments, it
is not sufficient to calculate the contraction with the metrical tensor
$-g_{\mu\nu}$ on the hadronic tensor. For our kinematical ranges at
 HERA (defined below), the ratio of ``$-g_{\mu\nu}H^{\mu\nu}$ / complete''
 is 0.75  (0.85)
for the ``low (high) $Q^2$ range''.
Leading order (LO)
 and next-to leading order  (NLO)
matrix elements for the processes $eP\rightarrow e + n \,\mbox{jets}
+ \mbox{remant jet} \hspace{4mm}
(n=1,2,3)$ are  implemented in \docuname.
Using an invariant jet definition scheme introduced in \cite{herai,heraii}
we present results for total NLO (2+1) jet cross sections
as well as differential distributions
in the basic kinematical variables $x, W^2$ and $Q^2$.
The dependence  on the renormalization scale $\mu_R$ and
factorization scale $\mu_F$ is investigated.
 It is shown that the NLO predictions
are much less sensitive to the choice of the scales that the LO results alone.
Varying $\mu^2=\mu_R^2=\mu_F^2$
in the range of $1/4 p_T^2$ to 4 $p_T^2$ induces an uncertainty of roughly
$\pm 3\%$ in the NLO (2+1) jet cross section predictions compared to
about $\pm 17 \%$ for the LO results for the
``high $Q^2$ range'' ($Q^2>100 \mbox{GeV}^2$) at HERA.
We have also studied the dependence on other choices of the scales
$[ \mu^2=0.1 Q^2 - 10 Q^2; y_{cut}W^2]$ and found that these variations lead to
larger uncertainties
(see table 1).
However, a choice of the scales like $\mu^2\sim p_T^2$
seems to be the  more appropriate choice in the case of (2+1) jet
production\footnote{An extreme example would be (2+1) jet production
at very low $Q^2$ ($Q^2\approx 1 \mbox{GeV}^2$),where  $Q^2$ cannot
be the relevant scale for high $p_T$ jet production. }.
We also explore the dependence of the (2+1) jet cross section on an additional
cut on the transverse momentum $p_T$ of the jets
($p_T$ is defined with respect  to the $\gamma^\ast$
direction).
 It is shown that most of  jets at a center of mass
(cm) energy of $\sqrt{s}=30$ GeV (E665
experiment) are produced with a transverse momentum less than 3 GeV using our
jet definition scheme.  Therefore, one should use
resummation techniques to obtain
 a reliable perturbation expansion in this region.
 This problem is not tackled in this paper.
Finally we also  present results for different values of the
 jet resolution parameter $y_{cut}$.
A comparison of jet measurements with our predicted QCD results provides a
direct tool of determining  \as\ or $\lms$.
Hadronization corrections may be minimized by restricting such an
analyses to large momentum transfer $Q^2$ which causes sufficiently large
transverse momenta of the participating partons.

The paper is organized as follows: In section II we discuss the general
structure of the cross sections used in the calculation. Section
III explains the jet definition scheme and in section IV we discuss
kinematical ranges and numerical results for (2+1) jet cross sections
 and rates. Finally, section V contains a short summary.

\section{Matrix elements}
Consider deep inelastic electron proton scattering
\begin{equation}
e^-(l) + \mbox{proton}(P) \rightarrow \mbox{proton remnant}(p_r) +
\mbox{parton} \,\,1 (p_1)
\ldots
+\mbox{parton}\,\, n (p_n)
\label{eq2}
\end{equation}
  Reaction (\ref{eq2}) proceeds via the exchange of an
intermediate vector boson $V=\gamma^\ast, Z,W$. In this paper only the
exchange by a virtual photon is considered.
We denote the
$\gamma^{\ast}$-momentum by $q$, the absolute square by $Q^2$, the
center of mass energy by s, the square of the final hadronic mass by
$W^2$ and introduce the scaling variables $x$ and $y$:
\begin{eqnarray} \label{defkin} q & = & l-l' \nonumber \\
Q^2 & \equiv & -q^2=xys>0 \nonumber \\
s & = & (P+l)^2 \nonumber \\ W^2 & \equiv & P_f^2=(P+q)^2 \\
x & = & \frac{Q^2}{2Pq} \hspace{1cm} (0<x \le 1) \nonumber \\
y & = & \frac{Pq}{Pl} \hspace{1cm} (0<y \le 1) \nonumber \end{eqnarray}
At fixed s, only two variables in (\ref{defkin}) are independent, since e.~g.
\begin{displaymath} xW^2=(1-x)Q^2,\hspace{1cm}  Q^2=xys. \end{displaymath}

When one of the hadronic final state momenta $p_1$
is measured, reaction (\ref{eq2}) is described in the one photon exchange
by five
parity conserved hadronic structure functions $H_1-H_5$ \cite{herai}
\begin{eqnarray}
H^{\mu\nu}&=& H_1 \left(g^{\mu\nu}-\frac{q^{\mu}q^{\nu}}{q^2}\right)
            + H_2  \frac{1}{Pq}\,\,\hat{P}^{\mu}\hat{P}^{\nu}
            + H_3  \frac{1}{Pq}\,\,\hat{p}_1^{\mu}\hat{p}_1^{\nu}\nonumber\\
        & + & H_4  \frac{1}{Pq}\,\,\left( \hat{P}^{\mu}\hat{p}_1^{\nu}
                        +\hat{P}^{\nu}\hat{p}_1^{\mu}\right)
            + H_5 \frac{1}{Pq}\,\, \left( \hat{P}^{\mu}\hat{p}_1^{\nu}
                        -\hat{P}^{\nu}\hat{p}_1^{\mu}\right)\label{hidef}
\end{eqnarray}
where we have introduced current conserved  momenta variables
$\hat{p}_i^{\mu}=p_i^{\mu}-\frac{(p_iq)}{q^2}q^{\mu}$.
The four $\mu\leftrightarrow\nu$ symmetric
  structure functions $H_1-H_4$  contribute to
so called T-even observables whereas the $\mu\leftrightarrow\nu$
antisymmetric structure function
 $H_5$ gives a contribution to
``T-odd'' observables \cite{hagiwara}.
To \oas \, in QCD one populates only $H_1-H_4$ as
there are no loop contributions to that order. LO contributions to
these structure functions have been extensively studied in the literature
\cite{herai,heraii,georgie,altarelli,koepp,mendez,kramer1,kramer2,soper}.

Note that in the totally inclusive case, where no final hadron momenta are
measured, one only has contributions to $H_1$ and $H_2$ which are then
denoted by the more familiar names $W_1$ and $W_2$.

Experimentally one can measure the so called helicity cross sections
$\sigma_{U+L}$, $\sigma_L$, $\sigma_T$, $\sigma_I$, $\sigma_A$ through
lepton hadron correlation effects.
They factorize the following $y$ and $\phi$ dependence \cite{herai}:
\begin{eqnarray}
d\sigma[n-\mbox{jet}]&\sim&\,\,\left[\,\,
(1+(1-y)^2)\,\,d\sigma_{U+L}[n-\mbox{jet}]\, -\,
y^2\,\,d\sigma_{L}[n-\mbox{jet}]\right.   \nonumber \\
&+&\,2(1-y)\cos 2\phi\,\,  d\sigma_{T}[n-\mbox{jet}]
-\,\,(2-y)\sqrt{1-y}\cos\phi\,\,d\sigma_{I}[n-\mbox{jet}]\nonumber\\
&+&\left.\,y\sqrt{1-y}\sin\phi\,\,d\sigma_{A}[n-\mbox{jet}] \right]
\label{angular}
\end{eqnarray}
In eq. (\ref{angular}) $\phi$ denotes
the  azimuthal angle between the parton plane
$(\vec{p},\vec{p_1})$ and the
lepton plane $(\vec{l},\vec{l'})$ (in the
($\gamma^\ast$-initial parton)-cms).

The helicity cross sections $\sigma_{X} \,\,(X\in\{U+L,L,T,I,A\})$
are linearly related to polarization density matrix elements
of the virtual $\gamma^\ast$. One has:
\begin{eqnarray}
 \sigma_{U+L} & \sim & h_{00}+h_{++}+h_{--}\label{hul}\\
 \sigma_{L} & \sim & h_{00} \hspace{4.45cm}\\
 \sigma_{T} & \sim & h_{+-}+h_{-+} \hspace{3.05cm}
            \\
 \sigma_{I} & \sim &  h_{+0} +h_{0+} -h_{-0}-h_{0-}
           \\
 \sigma_{A} & \sim &  h_{+0} +h_{0+} +h_{-0}+h_{0-}
\end{eqnarray}
where
$
h_{mm^{\prime}}= \epsilon_{\mu}^{\ast}(m)H^{\mu\nu}\epsilon_{\nu}(m^{\prime})
$, ($m,m^{\prime}=+,0,-$) and $\epsilon_{\mu}(\pm)
(\epsilon_{\mu}(0))$ are the
transversal (longitudinal) polarization vectors of the $\gamma^\ast$ in the
($\gamma^{\ast}-$initial parton)-cms\footnote{The
lepton hadron scattering process may be regarded as the scattering of a
polarized off-shell  gauge boson on the proton where the polarization of
the gauge boson is tuned by the scattered lepton's momentum direction.}.
Therefore   $\sigma_{U+L}$ and $\sigma_L$ labels the unpolarized
and longitudinal polarization,  $\sigma_T$ and $\sigma_I,\sigma_A$
mean transverse and transverse-longitudinal interference, respectively.
These helicity cross sections are also linearly related to  the five
covariant structure functions defined in eq. (\ref{hidef}) (see Apendix B of
\cite{herai}).

In this paper we present   results for (2+1)
jet cross sections where we have integrated over the azimuthal angle $\phi$.
Therefore only $\sigma_{U+L}$ and $\sigma_L$ contribute in
eq. (\ref{angular}).
These two cross sections can technically be obtained by the following
 covariant projections on the (partonic) hadrontensor $\hat{H}^{\mu\nu}$, which
 is calculated in fixed order perturbation theory ($p=\eta P$ denotes the
momentum of the incoming parton and $x_p=Q^2/(2pq)$):
\begin{eqnarray}
\sigma_{U+L}&=&\left(-\frac{1}{2}g_{\mu\nu}
               +\frac{3x_p}{pq}p_{\mu}p_{\nu}\right)
\,\,\hat{H}^{\mu\nu}[n-\mbox{jet}]\label{projections}\label{suldef}\\
\sigma_{L}&=&\frac{2x_p}{pq}p_{\mu}p_{\nu}\,\,\hat{H}^{\mu\nu}[n-\mbox{jet}]
\label{sldef}
\end{eqnarray}

The \oasz\ (2+1) jet  matrix elements represent a full NLO
 calculation including virtual and real corrections.
The following subprocessess contribute to (2+1) jet cross sections up to \oasz:
\begin{eqnarray}
\hat{H}^{\mu\nu}[\mbox{tree},\oas \,]&:&   \gamma^\ast+q \rightarrow q
                                              + G \nonumber\\
                     &&           \gamma^\ast+G \rightarrow q + \bar{q}
                                               \nonumber\\
\hat{H}^{\mu\nu}[\mbox{tree},\oasz \,]&:&  \gamma^\ast+q \rightarrow q + G + G
                                                \nonumber\\
                     &&           \gamma^\ast+q \rightarrow q + \bar{q} + q
                                                \label{subproc}\\
                     &&           \gamma^\ast+G \rightarrow q + \bar{q} + G
                                                 \nonumber\\
\hat{H}^{\mu\nu}[\mbox{virtual},\oasz \,]&:&\gamma^\ast+q \rightarrow q + G
                                                \nonumber\\
                     &&           \gamma^\ast+G \rightarrow q + \bar{q}
                                    \nonumber
\end{eqnarray}
and the corresponding anti-quark processes with $q\leftrightarrow \bar{q}$.
  Matrix elements for the
{\it complete} contributions to (2+1) jet NLO $O(\alpha_s^2)$
corrections are first discussed in \cite{tomzpc}.
The NLO matrix elements used in the
MC program \docuname\ are based
on these matrix elements. In addition, we have also
included the full NLO scale dependent
contributions.
Note, that the second projection ($\sim p_{\mu}p_{\nu}$) in $\sigma_{U+L}$
in eq. (\ref{suldef}) gives a contribution of the order of
$15-30\%$ to the (2+1) jet cross sections depending on the kinematical ranges,
wheras the contribution from $\sigma_L$ in eq. (\ref{sldef}) is fairly small
(less than 1\% in our kinematical ranges). This originates from the
$y$ dependent coefficients $(1+(1-y)^2)$ and $-y^2$ and the fact that $y$
is peaked at small values (see  fig. 9 in \cite{heraii}).
The  (2+1) jet NLO
contributions originating from  the  projection with $-g_{\mu\nu}$ on the
hadron tensor (see eq. (\ref{projections})) are first presented and discussed
in detail
in \cite{dirk}. A complete list of tree level matrix elements with up to
four partons  in the final state can also be found in \cite{dieter}.

The general structure of the NLO jet   cross sections in DIS within the
framework of perturbative QCD in given by:
\begin{equation}
d\sigma^{had}[(2+1)-\mbox{jet}] =
\int d\eta \,\,f_a(\eta,\mu_F^2)\,\,\, d\hat{\sigma}^a(p=\eta P,
\alpha_s^2(\mu_R^2), \mu_R^2, \mu_F^2)
\end{equation}
where one sums over $a=q,\bar{q},g$. $f_a(\eta,\mu_F^2)$
is the probability density to find a parton $a$ with fraction $\eta$
in the proton if the proton is probed at a scale $\mu_F^2$.
$\hat{\sigma}^a$ denotes the partonic cross section
from which collinear initial state
singularities have been factorized out
at a scale $\mu_F^2$ and
implicitly included in the scale dependent parton densities
$f_a(\eta,\mu_F^2)$.

 Let us briefly discuss some
technical matters that go into the NLO calculation.
The \oasz \, tree graph contributions in eq. (\ref{subproc})
are integrated over the unresolved phase space region which are (2+1)
jet like as defined in eq. (\ref{jetdef}).
Infrared (IR) as well as collinear (M) divergencies associated
with the final state partons are cancelled against corresponding IR/M
divergencies of the one-loop contributions.
The remaining collinear initial state divergencies are factorized into
the bare parton densities introducing a factorization scale dependence
$\mu_F$. Finally the ultraviolet (UV) divergencies are removed by
$\overline{MS}$ renormalization
 which introduces a renormalization scale dependence
$\mu_R$.

\section{Jet definition}
In order to calculate the (2+1) jet cross section we have to {\em define}
what we call (2+1) jets by introducing a {\em resolution criterion}.
As has been elaborated in detail in \cite{herai,heraii}
energy-angle cuts are
not suitable for an asymmetric machine with its strong
boosts from the hadronic cms to the laboratory frame.
As a jet resolution criterion we use the invariant mass cut
criterion  defined in\cite{herai,heraii,tomzpc,dirk}
such that
\begin{equation}
s_{ij}=(p_i+p_j)^2 \geq M^2=\max\left\{ y_{cut} M_c^2,M_0^2\right\}
\hspace{1cm}
(i,j=1,\ldots ,n,r;\hspace{5mm} i\neq j)
\label{jetdef}
\end{equation}
where $y_{cut}$ is the resolution parameter  and $s_{ij}$
is the invariant mass of any two final state partons,
including the remnant jet with momentum
$ p_r=(1- \eta )P $.
 $M_c$ is a
typical mass scale of the process which defines the jet definition scheme.
In this paper we choose $M_c^2=W^2$.
This corresponds to the ''$W$-''scheme in \cite{herai}. Other jet definition
schemes based on a $k_T$ algorithm are proposed in \cite{catani}.
$M_0$ is a fixed mass cut which we have introduced
in order to clearly separate the perturbative and non-perturbative
regime in the case where $W^2$ is small.
$M_0$  is  fixed to 2 GeV  \cite{heraii} in all our results.\\

\section{Numerical results}
We will now turn to our numerical cross section results
and  present results  for the actual  HERA cm energy of 295 GeV as well as for
the
FERMILAB fixed target experiment E665 with a cm energy of 30 GeV.
All results are based on a new Monte Carlo program \docuname\ \cite{disjet}.
The Monte-Carlo routines are using the VEGAS-package \cite{vegas}
for numerical integration. Parton distributions are incorporated from the
packages \cite{pakpdf,pdflib}.
Our standard set of parton distribution functions is MRS set
D- \cite{mrs}. If not stated otherwise,
 we
use the one-loop formula for the strong coupling constant \as\, for our LO
results
\label{aslo}
\begin{equation}
   \alpha_{s\,\overline{MS}}(\mu_R^2)
= \frac{12\pi}{(33-2n_f)\,\ln\frac{\mu_R^2}{\Lambda^2}}
\end{equation}
whereas we employ the two loop formula
\begin{equation}
\alpha_{s\,\overline{MS}}(\mu_R^2)=
\frac{12\pi}{(33-2n_{f})\ln(\mu_R^2/\Lambda^{2})}
\left[1-\frac{6(153-19n_{f})}{(33-2n_{f})^{2}}
\frac{\ln\ln(\mu_R^2/\Lambda^{2})}{\ln(\mu_R^2/\Lambda^{2})}\right]
\label{asnlo}
\end{equation}
in our NLO predictions.
The $\lms$ value is chosen  consistent to the $\Lms{4}$ value from the
parton distribution functions.
The value of $\alpha_s$ is matched at the thresholds $q=m_q$ and the
number of flavours $n_f$ in $\alpha_s$ is fixed by the number
of flavours for which the masses are less than $\mu_R$.
Furthermore the number of quark flavours that can be pair-produced
are set equal to $n_f$ chosen in $\alpha_s$.
In all predictions, the renormalization scale  and
the factorization scale  are set to be equal: $\mu_R^2=\mu_F^2=\mu^2$.
For the fine structure constant $\alpha$ we adopt the running coupling formula.

The following kinematical cuts are used for the HERA  results:
\begin{equation}
\label{rangehera}
\matrix{
0.001              &< & x    &  <  &  0.1  \cr
0.04               &< & y    &  <  &  0.95 \cr
600\,\, \mbox{GeV}^2   &< & W^2  &     &       \cr
}
\end{equation}
In addition, we use two different $Q^2$ ranges:\\
''Low $Q^2$ range:''
 $4 \,\,\mbox{GeV}^2  \,\,< \,\, Q^2 \,\,< 100 \,\, \mbox{GeV}^2$\\
''high $Q^2$ range:'' $100\,\, \mbox{GeV}^2  \,\,< \,\, Q^2$\\
Note that the kinematical cuts in eq. (\ref{rangehera})
are not independent, for example
the $x$ and $y$ cut also imply  $W>55$ GeV ($W^2=(1-x)ys$) and  $x_{max}$
in the low $Q^2$ range is $x_{max}=0.0287$ rather than 0.1 ($Q^2=xys$).

Our kinematical range for the E665 experiment is defined by \cite{e665pub}:
\begin{equation}
\label{rangee665}
\matrix{
0.003              &< & x    &                          \cr
0.08               &< & y    &  <  &  0.95              \cr
4 \,\,\mbox{GeV}^2     &< & Q^2  &  <  &  25\,\, \mbox{GeV}^2   \cr
400 \,\,\mbox{GeV}^2    &< & W^2  &     &                    \cr
}
\end{equation}

In fig.~1 we show the dependence of the (2+1) jet rate on the resolution
parameter $y_{cut}$ for the two different $Q^2$ ranges at HERA (a,b)
and for the E665 experiment (c). Solid (dashed) lines correspond to
NLO (LO) predictions.
The renormalization scale $\mu_R^2$ and the
 factorization scales $\mu_F^2$ are set equal to $Q^2$
(lower curves) and $p_T^2$ (upper curves).
The transverse momentum $p_T$ of the jets  is defined in eq. (\ref{pt2def}).
One observes, that the NLO corrections lower the
(2+1) jet rate by about 15\%.
At $y_{cut}=0.02$
 the (2+1) jet rate
exceeds about 4\% (12\%) for the low (high) $Q^2$
range at HERA.
For the E665 experiment,
the (2+1) jet rate exceeds about 10-11\%  for $y_{cut}=0.04$.
In the following, we use $y_{cut}=0.02$ for HERA and
$y_{cut}=0.04$ for E665 as our standard values.

To get a feeling for the theoretical uncertainties originating from
the choice of the renormalization and factorization scales, we show
numerical results for the (2+1) jet cross sections for different $\mu^2$
values in table 1. Let us first comment on the choice of the scales.
In DIS scattering it is natural to take
$\mu^2=\mu_R^2=\mu_F^2=Q^2$.  However, for  jet production,
the transverse momentum $p_T$ of the jets
 should  also be considered as a relevant scale.
The transverse momentum
 is defined with respect to the $\gamma^\ast$ direction.
 For the (1+1) case, one has two  jets,
the remant jet and the
struck parton
 jet, both with zero $p_T$\footnote{ The intrinsic transverse momentum
of the partons in the target is neglected}.
 In our case of (2+1) jet production, one expects
two partonic jets with
nearly opposite large $p_T$ and the remnant jet at $p_T=0$.
For the LO processes, $p_T$ is given by:
\begin{equation}
p_T^2= Q^2\frac{1-x_p}{x_p}z(1-z)
\label{pt2def}
\end{equation}
with
\begin{equation}
x_p=\frac{Q^2}{2pq}=\frac{x}{\eta} \hspace{1cm} z=\frac{pp_1}{pq}
\end{equation}
where $p_1$ is the four momentum of one of the final partons.
Finally, in analogy to the jet analysis in $e^+e^-$ experiments
 one may use $y_{cut} W^2$ (see. eq. (\ref{jetdef})) as a possible scale.
To avoid to small
scales for perturbation theory, the
scales are clipped at a minimum value of 2 GeV$^2$.
Varying $\mu^2=\mu_R^2=\mu_F^2$  between $ Q^2$,\,
 $0.25 \,p_T^2  - 4\, p_T^2$  and $y_{cut} \,W^2$ induces
 an uncertainty of roughly
$\pm$ 17\% ($\pm$ 4\%) in LO (NLO) for the high $Q^2$ range at HERA
 and
$\pm$ 20\% ($\pm$ 10\%) in LO (NLO) for the E665 experiment.
Therefore, the uncertainty in the theoretical predictions is
markedly reduced by the NLO corrections.

In figs. 2-10 we show the dependences of the total cross section and the (2+1)
jet cross sections as well as the (2+1) jet rate on the basic kinematical
variables $x, W^2$ and $Q^2$.
Figs. a (b) are for the low (high) $Q^2$
range at HERA and c shows predictions for the E665 experiment.
LO results are given in figs. 2,3,6,7,11-15 whereas figs. 4,5,8,9 and 10
show NLO predictions.

In order to estimate the theoretical uncertainty from the
choice of the  scales,
all results are given for
$\mu^2=\mu_R^2=\mu_F^2= a \,p_T^2\,\,\,(a$=1/4,\,1,\,4: solid lines),
$\mu^2= a \,Q^2\,\,\,(a$=1/10,\,1,\,10: dashed lines) and
$\mu^2=y_{cut} W^2$ (dotted lines).

Let us first comment on the $W$ distributions shown in figs. 2-5.
The (2+1) jet rate decreases with increasing $W$. This is mainly
a reflection of our jet definition in eq. (\ref{jetdef}) where the
required invariant mass of the jets increases with increasing $W^2$.
The (2+1) jet rate for the high $Q^2$ sample is larger
than for the low $Q^2$ range since the larger $Q^2$ causes at the average
larger transverse momenta of the participating partons
(see  eq. (\ref{pt2def})) and  larger invariant masses.
Comparing figs. 2,3 and figs. 4,5
one observes again that the uncertainty from the choice of the scales in the LO
predictions is very large (figs. 2,3) whereas this
uncertainty is markedly reduced by
including the NLO corrections (figs 4,5).

Turning now to the $x$ distributions one observes a quite different
behaviour of the cross sections for the different kinematical ranges
in figs.  6 and 8.
The $x$ distributions are mainly governed by the behaviour of the
respective parton distributions in the allowed kinematical regions.
Note that $x_{max}=0.0287$ for the low $Q^2$ range (figs 6a and 8a).
The
(2+1) jet rates in figs. 7 and 9 are increasing with increasing $x$.
This is mainly an effect of the increasing  (2+1) jet
 phase space in our jet definition scheme
(for a detailed discussion of the $\oas$  (2+1) jet phase space
see \cite{herai}).
Comparing the
distributions in figs. 6-9, one observes again that
the NLO predictions in figs. 8,9 are much less scale dependent than the
LO predictions in figs. 6,7.

Fig. 10 shows NLO predictions for the (2+1) jet rate as a function
of $\sqrt{Q^2}$. Results are given for $\alpha_s(\mu_R^2=Q^2)$ (lower solid
curve)
$\alpha_s(\mu_R^2=p_T^2)$,
 (upper solid curve) and $\alpha_s=\mbox{fix}=0.25$ (dotted
curve).
Note that the (2+1) jet rate is an increasing function with increasing
$Q^2$. Therefore the $Q^2$ or $p_T^2$ dependence in $\alpha_s$ is
 overcompensated  by the increasing  (2+1) jet phase space
 relative to the total cross section in our
jet definition scheme.
One observes sizable differences between the predictions using  a constant
$\alpha_s$ and scale dependent (two-loop) $\alpha_s$.
Therefore a clear discrimination
between the solid and dotted curves should be possible with the
expected event statistics
at HERA.

Note also, that  $\mu^2=p_T^2$
tends to  predict larger (2+1) jet rates and differential distributions
 than $\mu^2=Q^2$
 (see also table 1 and figs. 1-10).
This is in particular true for lower $W^2$ and
higher $x$ values.

In order to explore the $p_T$ dependence of the jet rates in more detail, fig.
11 shows results for the (2+1) jet rate as function of an additional $p_T$ cut.
Applying a $p_T$ cut of 4 GeV reduces the (2+1) jet rate for the low (high)
$Q^2$ range at HERA from $\sim 4$\% to
2-3\% ($\sim 12$\% to $\sim 10$\%) whereas the (2+1)
jet rate falls from 10\% to less than  1\% for the FERMILAB experiment.
Fig. 11c shows also, that more than 50\% of the jets at
$\sqrt{s}=30$ GeV  are produced
with a  transversal momentum less than 2 GeV.
However, these values are too small to allow for reliable predictions
in fixed order perturbation theory and resummation techniques should
be used in this kinematical region.
In fact , the experimental results presented by the E665
collaboration \cite{e665pub}
are significantly higher than our second order predictions in figs. 1-9 c.

In table 2, we give results for jet cross sections with an additional
$p_{T\,min}$ cut. Figs. 12-15 show the corresponding $W$ and $x$ distributions.
Note that the  $(2+1)$ jet rates  in figs. 12-15
are less $x$ and $W$ dependent than our results without the additional
$p_{T\,min}$ cut. This
demonstrates that the increase of the jet rates for small $W$ (figs. 3,5)
and large $x$ values (figs. 7,9) is mainly an effect of low $p_T$ jets.
This is also clear by an inspection of eq. (\ref{pt2def}).


\section{Summary}
The (2+1) jet production in DIS is calculated up to NLO in
perturbative QCD using an invariant jet definition scheme.
 The theoretical uncertainties originating from variations of the
renormalization/factorization scales
are well under control.  Differential distributions of
(2+1)
jet rates in $W^2,x$ and $Q^2$ are presented. A comparison with the
large numer of events expected at HERA in the near future
will allow for precision tests of  perturbative
QCD.
It is shown that an additional $p_T$ cut to our jet definition
scheme is necessary to obtain reliable predictions for jet production
at the energy of the
E665 experiment at FERMILAB.\\[1cm]
\noindent
{\bf Acknowledgements:}\\
We thank J.G. K\"orner for interesting collaboration in part of this work.
We thank I.H. Park
for his efforts to test the MC program \docuname\
 and helpful suggestions concerning
some extensions of the program.
E.M.  thanks C. Salgado and
the ZEUS groups at Madison and Argonne for very  stimulating discussion.
This work is supported in part by the U.S. Department of Energy under
contract No. DE-AC02-76ER00881, and in part by the University of Wisconsin
Research Committee with funds granted by the Wisconsin Alumni Research
Foundation.

\newpage
\def\npb#1#2#3{{\it Nucl. Phys. }{\bf B #1} (#2) #3}
\def\plb#1#2#3{{\it Phys. Lett. }{\bf B #1} (#2) #3}
\def\prd#1#2#3{{\it Phys. Rev. }{\bf D #1} (#2) #3}
\def\prl#1#2#3{{\it Phys. Rev. Lett. }{\bf #1} (#2) #3}
\def\prc#1#2#3{{\it Phys. Reports }{\bf C #1} (#2) #3}
\def\pr#1#2#3{{\it Phys. Reports }{\bf #1} (#2) #3}
\def\zpc#1#2#3{{\it Z. Phys. }{\bf C #1} (#2) #3}
\def\ptp#1#2#3{{\it Prog.~Theor.~Phys.~}{\bf #1} (#2) #3}
\def\nca#1#2#3{{\it Nouvo~Cim.~}{\bf #1A} (#2) #3}
%

\begin{table}
\renewcommand{\baselinestretch}{1.6}
\begin{center}
\begin{tabular}{|r|c|c|c|c||c|c|}
$\mu^2 \hspace{2mm} $    & low $Q^2$: LO    &  low $Q^2$: NLO        &
 high $Q^2$: LO       & high $Q^2$: NLO  & E665: LO    & E665: NLO   \\
\hline \hline
$Q^2\hspace{2mm} $         & 2360 & 1930 & 600 & 547  & 1200   & 1020
\\   \hline
$10\cdot Q^2\hspace{2mm} $ & 1980 & 1930 & 472 & 502  & 879    & 932
\\   \hline
$0.1\cdot Q^2\hspace{2mm}$ & 3050 & 2000 & 770 & 548  & 1660   & 1130
\\   \hline
$y_{cut}W^2\hspace{2mm} $  & 2080 & 1960 & 602 & 553  & 1040   & 983
\\   \hline
$p_T^2\hspace{2mm}$        & 2320 & 2020 & 719 & 579  & 1310   & 1120
\\   \hline
$4\cdot p_T^2\hspace{2mm}$ & 2100 & 2000 & 615 & 562  & 1040   & 1010
\\   \hline
\hspace{2mm}$0.25\cdot p_T^2\hspace{2mm} $
                           & 2670 & 2140 & 857 & 583  & 1590   & 1180
\end{tabular}
\renewcommand{\baselinestretch}{1.1}
\normalsize
\end{center}
\caption{
(2+1) jet cross sections in [pb]
 for different choices of the renormalization and
factorization scales (column 1). LO and NLO results are shown for
the ``low $Q^2$'' (column 2 and 3) and the ``high $Q^2$''
(column 4 and 5) ranges at HERA ($y_{cut}=0.02$)
 and for the E665 experiment (column 6 and 7) ($y_{cut}=0.04$).
 \label{tab1}
}
\end{table}
\begin{table}
\renewcommand{\baselinestretch}{1.6}
\begin{center}
\begin{tabular}{|r|l|l|l|}
$\mu^2 \hspace{2mm} $   & low $Q^2$  & high $Q^2$  & E665    \\
\hline \hline
$p_T^2\hspace{2mm}$        & 1140    & 451    &  75 \\   \hline
$4\cdot p_T^2\hspace{2mm}$ & 1040    & 394    &  59\\   \hline
\hspace{2mm}$0.25\cdot p_T^2\hspace{2mm} $
                           & 1230    & 514    & 103
  \end{tabular}
\renewcommand{\baselinestretch}{1.1}
\normalsize
\end{center}
\caption{
LO predictions for (2+1) jet cross sections in [pb] with an additional
$p_{T\,min}$ cut of 4 GeV. The two-loop formula for $\alpha_s$ is used
in the calculation.
 \label{tab2}
}
\end{table}

\newpage
\section*{Figure Captions}
\begin{enumerate}
\item[Fig.~1.]
(2+1) jet fraction versus the cut variable $y_{cut}$ for the low and high
$Q^2$ range at HERA (a,b) and for the E665 experiment (c).
 The kinematical ranges are defined in eqs.
(\ref{rangehera},\ref{rangee665}).  Solid (dashed) lines correspond to
NLO (LO) predictions.
In the   upper (lower) lines
the renormalization scale
$\mu_R$ and
 factorization scale $\mu_F$ are set equal to
$\mu^2=p_T^2$ ($\mu^2=Q^2$).
\item[Fig.~2.]
$W$ dependence of the total cross section (upper line) and
the (2+1) jet cross section in LO for the
low $Q^2$ (a) and high $Q^2$ (b) range at HERA and for the E665
 experiment (c).
$y_{cut}=0.02 \,\,(0.04)$ in a,b (c).
The different curves for the jet
cross section belong to different choices of the renormalization scale
$\mu_R$ and
 factorization scale $\mu_F$:\\
 solid lines: $\mu^2=\mu_R^2=\mu_F^2= 1/4 p_T^2,\, p_T^2,\, 4 p_T^2$
(from top to bottom).\\
 dashed lines: $\mu^2= 0.1 Q^2, Q^2, 10 Q^2$
(from top to bottom).\\
 dotted line: $\mu^2= y_{cut} W^2$.

\item[Fig.~3.]
$W$ dependence of the
(2+1) jet rate in LO.
Solid, dashed and dotted lines as in fig. 2.

\item[Fig.~4.]
same as fig.~2, but NLO predictions.
The two upper lines in c) show the LO (dashed) and NLO ($O(\alpha_s)$, solid)
predictions for the total cross section. The difference between these results
for HERA energies are too small to be visible in a) and b).
\item[Fig.~5.]
same as fig.~3, but NLO predictions.

\item[Fig.~6.]
$x$ dependence of the total cross section (upper line) and
the (2+1) jet cross section in LO for the
low $Q^2$ (a) and high $Q^2$ (b) range at HERA and for the E665
experiment (c).
Solid, dashed and dotted curves as in fig.~2.

\item[Fig.~7.]
$x$ dependence of the (2+1) jet rate in LO.
Solid, dashed and dotted curves as in fig.~6.

\item[Fig.~8.]
same as fig. 6, but NLO predictions.

\item[Fig.~9.]
same as fig. 7, but NLO predictions.

\item[Fig.~10.]
$\sqrt{Q^2}$ dependence of the (2+1) jet rate in NLO for $\mu^2=Q^2$
 (lower solid line) $\mu^2=p_T^2$ (upper solid line).
Also shown is the result for $\alpha_s= \mbox{const.} = 0.25$ (dotted line).
\item[Fig.~11.]
LO (2+1) jet rate as a function of $p_{T\, min}$ for the low and high
$Q^2$ range at HERA with $y_{cut}=0.02$ (a,b) and for E665
 experiment (c) $(y_{cut}=0.04)$. The three curves are for
$\mu^2=\mu_R^2=\mu_F^2= 1/4 p_T^2,\, p_T^2,\, 4 p_T^2$ (from top to bottom).
We have used the two-loop formula for $\alpha_s$.

\item[Fig.~12]
$W$ dependence of the total cross section (upper line) and
the (2+1) jet cross section in LO for the
low $Q^2$ (a) and high $Q^2$ (b) range at HERA and for the E665 experiment (c)
with an additional $p_{T\, min}$ cut of 4 GeV (for the  jet cross sections).
Same parameters as in fig.~11 for the (2+1) jet cross sections.

\item[Fig.~13]
$W$ dependence of the
(2+1) jet rate in LO with  an  $p_{T\, min}$ cut of 4 GeV
(parameters as in fig. 11).

\item[Fig.~14.]
$x$ dependence of the total cross section (upper line) and
the (2+1) jet cross section in LO for the
low $Q^2$ (a) and high $Q^2$ (b) range at HERA and for the E665 experiment(c)
(parameters as in fig. 11).

\item[Fig.~15.]
$x$ dependence of the
(2+1) jet rate in LO with  an  $p_{T\, min}$ cut of 4 GeV
(Parameters as in fig. 11).
\end{enumerate}


\begin{thebibliography}{99}
 \bibitem{h1pub} H1 Collaboration, \plb{299}{1993}{385};\plb{298}{1993}{469}
 \bibitem{zeuspub} ZEUS Collaboration, \plb{306}{1993}{158};
                   DESY preprint 93-030
 \bibitem{e665pub} E665 Collab.,  \plb{272}{1991}{163}\\
                   E665 Collab.,  \prl{69}{1992}{1026}\\
                   E665 Collab., FERMILAB-PUB-93/171-E\\
                   C. Salgado., FERMILAB-Conf-93/251-E
 \bibitem{disjet} T. Brodkorb and E. Mirkes, Madison preprint, MAD/PH/821.
 \bibitem{herai} J.G. K\"orner, E. Mirkes, G. Schuler: {\it Int. J. Mod. Phys.}
                    {\bf A4} {1989} 1781
 \bibitem{hagiwara} K. Hagiwara, K. Hikasa and N. Kai, \prd{27}{1983}{84}
 \bibitem{heraii} T. Brodkorb, J.G. K\"orner, E. Mirkes
                  G. Schuler: \zpc{44}{1989}{415-432}
 \bibitem{georgie} H. Georgie and H.D. Politzer, \prl{40}{1978}{3}
 \bibitem{altarelli} G. Altarelli and B. Martinelli \plb{76}{1978}{89}
 \bibitem{koepp}   G. K\"opp, R. Maciejko and P.M. Zerwas \npb{144}{1978}{123}
 \bibitem{mendez}  A. Mendez,  \npb{145}{1978}{199}
 \bibitem{kramer1} G. Kramer and C. Rumpf,  \plb{89}{1980}{380}
 \bibitem{kramer2} G. Kramer and C. Rumpf and J. Willrodt,
                       \zpc{7}{1981}{337}
 \bibitem{soper}   R. Meng, F.I. Olness and D.E. Soper, \npb{371}{1992}{79}
 \bibitem{tomzpc} T. Brodkorb, J.G. K\"orner: \zpc{54}{1992}{519}.
 \bibitem{dirk} D. Graudenz \plb{256}{1992}{518}, preprint LBL-34147 (1993)
 \bibitem{dieter} K. Hagiwara and D. Zeppenfeld, \npb{313}{1989}{39}.
 \bibitem{catani} S. Catani, Yu.L. Dokshitzer and B. Webber,
                    \plb{285}{1992}{291}
 \bibitem{vegas} G. P. Lepage, {\it J. Comp. Phys.} {\bf 27} (1978) 192.
 \bibitem{pakpdf} K. Charchula, {\it Comp. Phys. Comm.} {\bf 69} (1992) 360
 \bibitem{pdflib} H. Plothow-Besch,  {\it CERN-PPE}/92-123
 \bibitem{mrs}  A.D. Martin, W.J. Stirling ans R.G. Roberts,
                  \prd{47}{1993}{867}
\end{thebibliography}
\end{document}